\documentclass[prologue,table,xcdraw,screen,nonacm]{acmart}
\usepackage{enumitem}

\AtBeginDocument{%
  \providecommand\BibTeX{{%
    \normalfont B\kern-0.5em{\scshape i\kern-0.25em b}\kern-0.8em\TeX}}}

\acmBooktitle{Twenty-ninth ACM SIGKDD Conference (KDD '23), August 6--10, 2023, Long Beach, US}

\begin{document}

\title{EvalRS 2023. Well-Rounded Recommender Systems For Real-World Deployments}



\author{Federico Bianchi}
\affiliation{%
  \institution{Stanford}
  \city{Palo Alto}
  \country{USA}}
\email{fede@stanford.edu}

\author{Patrick John Chia}
\affiliation{%
  \institution{Coveo}
  \country{Canada}
}

\author{Ciro Greco}
\email{ciro.greco@bauplanlabs.com}
\affiliation{%
  \institution{Bauplan}
  \country{USA}
}

\author{Claudio Pomo}
\affiliation{%
  \institution{Politecnico di Bari}
  \country{Italy}
}

\author{Gabriel Moreira}
\affiliation{%
  \institution{NVIDIA}
  \streetaddress{30 Shuangqing Rd}
  \city{Sao Paulo}
  \state{Sao Paulo}
  \country{Brazil}}

\author{Davide Eynard}
\affiliation{%
  \institution{mozilla.ai}
  \country{United Kingdom}
}
\email{davide@mozilla.ai}

\author{Fahd Husain}
\affiliation{%
  \institution{mozilla.ai}
  \country{Canada}
}
\email{fahd@mozilla.ai}

\author{Jacopo Tagliabue}
\affiliation{%
 \institution{Bauplan, New York University}
 \country{USA}}

\renewcommand{\shortauthors}{Bianchi et al.}



\begin{abstract}

EvalRS aims to bring together practitioners from industry and academia to foster a debate on rounded evaluation of recommender systems, with a focus on real-world impact across a multitude of deployment scenarios. Recommender systems are often evaluated only through accuracy metrics, which fall short of fully characterizing their generalization capabilities and miss important aspects, such as fairness, bias, usefulness, informativeness. This workshop builds on the success of last year's workshop at CIKM, but with a broader scope and an interactive format.

\end{abstract}




\maketitle


\section{Introduction}
The ubiquity of personalized recommendations in various online platforms, from e-commerce to news to social media, has led to a surge of interest in recommender systems (RS) research. The field has grown to accommodate the new application scenarios, resulting in a plethora of algorithmic approaches to address modelling challenges~\cite{DBLP:journals/aim/JannachPRZ21,DBLP:conf/recsys/Tagliabue21}. However, evaluating RS performance is still difficult, especially considering the increasing complexity of deployments and variety of use cases~\cite{DBLP:conf/recsys/JannachMO20,DBLP:conf/recsys/HigleyOARM22}. In this light, our workshop focuses on multi-dimensional and multi-faceted evaluation techniques: while accuracy metrics are often seen as the proxy for generalization, they miss other important dimensions of real-world systems, such as fairness, informativeness, transparency, and resource constraints~\cite{DBLP:conf/www/ChiaTBHK22}.

By expanding evaluation beyond traditional accuracy metrics, we aim to better understand the holistic performance of RS across diverse scenarios. The workshop gives participants an in-depth view of multi-dimensional evaluation techniques, allowing them to acquire fundamental skills as well as ``live and breath'' the problem through a novel format, our \textit{hackathon} (Section~\ref{sec:hackathon}). 

As the enthusiasm about the first edition (EvalRS 2022) showed appetite in the community for re-assessing testing practices, we believe it is time for a new, revised, and improved version of EvalRS. As highlighted even in a recent editorial piece in \textit{Nature Machine Intelligence}, RS ``benefits can also give rise to challenging ethical issues'' \cite{natureMIchallenge}: it is clear that both researchers and industry practitioners want to better understand how to avoid negative societal consequences, such as unfair treatment of users, harmful echo chambers, bias, and increasing levels of polarization and misinformation.

A more nuanced approach to evaluation is essential against this backdrop. With the help of new organizers and the support of mozilla.ai, EvalRS 2023 will help bring these important themes front and center within our community and spark a lively debate on the oldest question of all: how do we know we are doing the right thing?

\section{Workshop Description}

EvalRS has set out to accomplish a significant goal: to foster closer partnerships between the academic and industrial sectors regarding RS. This is achieved by placing a strong emphasis on comprehensive evaluation techniques that can be effectively applied across diverse domains where RS are utilized. The traditional approach to RS evaluation has been centered on accuracy metrics~\cite{DBLP:conf/naacl/BianchiTY21,DBLP:conf/recsys/KoukiFVCLJ20,DBLP:conf/recsys/RashedJSH20,DBLP:conf/recsys/SunY00Q0G20,DBLP:conf/um/AnelliBNJP22}. However, EvalRS seeks to expand the scope of evaluation techniques beyond just accuracy, to encompass other vital aspects such as fairness, interpretability, and robustness. The ultimate aim of EvalRS is to promote transparency and accountability in the development and deployment of RS by encouraging ethical evaluation metrics that prioritize end-users' interests.

This year's workshop aims to expand on last year's successful event at CIKM. EvalRS2022 was organized as an open source competition (>150 participants in 50 teams) and a popular half-day workshop (>50 attendees). All the materials from the workshop have been released to the public, including accepted papers, models, tests, tutorials, and keynote.\footnote{See \url{https://github.com/RecList/evalRS-CIKM-2022/blob/main/README_CIKM_2022.md} for all the links.} The event was a first-of-its-kind initiative and a retrospective has been recently published in Nature Machine Intelligence~\cite{tagliabue2023challenge}.

Building on the experience and expertise of our diverse group of organizers and PC members, this upcoming edition of our program will be even more comprehensive than before. We are committed to maintaining the open-source spirit that was a hallmark of the first edition while introducing a novel interactive element - a hackathon on evaluation - to make this workshop even more distinctive.

The hackathon will provide participants with a unique opportunity to put their evaluation skills to the test and collaborate with others in real-time. Participants will have the opportunity to work together to solve practical evaluation challenges and gain valuable hands-on experience in the process. The hackathon will serve as an excellent platform for networking and sharing knowledge with other experts in the field.

\subsection{Workshop Theme}

EvalRS aims to foster a debate on the rounded evaluation of RS~\cite{DBLP:journals/aim/CremonesiJ21, DBLP:journals/aim/JannachPRZ21}, with a focus on real-world impact across a multitude of deployment scenarios. By bringing together experts from industry, academia, and government, EvalRS creates a forum for discussion and collaboration on the latest trends and challenges across a wide range of domains. The main themes from EvalRS 2022 -- slice-based metrics, fairness assessment, and the use of representational learning to scale behavioral tests -- will be expanded upon in this edition, with particular attention to the social impact of RS~\cite{DBLP:journals/ais/MilanoTF20,DBLP:journals/ipm/ElahiKKSRT21}. For these reasons, we expect EvalRS 2023 to attract a broad range of practitioners, reflecting the horizontal nature of evaluation challenges: taking last year as an example, participant affiliations ranged from academia (e.g. Politecnico di Milano) to startups (Coveo), from Big Tech (Microsoft) to traditional corporations (Fidelity Investment).

The importance of developing better testing for any deployed system can hardly be overstated, let alone for systems as ubiquitous as RS.
We believe that the rounded evaluation of RS is, by nature, a multi-faceted and multi-disciplinary endeavor and that the field as a whole has often been held back by the false dichotomy of \textit{quantitative-and-scalable} vs \textit{qualitative-and-manual}~\cite{DBLP:conf/www/ChiaTBHK22}. This workshop can potentially drive innovation and advancement in IR and adjacent fields through our commitment and experience in bridging the gap between industry and academia. 

\subsection{Call for Papers}
\label{sec:CFP}

We encourage the submission of original contributions along our main topics. Submitted papers will be evaluated (single-blind) according to their originality, technical content, style, clarity, and relevance to the workshop. Papers must be original work and may not be under submission to another venue at the time of review. Accepted papers will appear in the \textbf{workshop proceedings}\footnote{We plan on publishing on CEUR, as we did for EvalRS 2022.}.

As a non-exhaustive list, we encourage submissions of long research and position papers (up to 8 pages), short research and position papers (up to 4 pages + refs) and long abstracts (up to 2 pages + refs) on the following topics:

\begin{itemize}
\item Online vs offline evaluation - e.g. making offline evaluation more trustworthy and unbiased;
\item Tools and frameworks for the evaluation of RS;
\item Empirical studies on the evaluation of RS;
\item Reports from real-world deployments - failures, successes, and surprises;
\item New metrics and methodologies for evaluation, both quantitative and qualitative;
\item Multi-dimensional evaluation, combining multiple recommendation quality factors;
\item Multi-disciplinary investigation on ethical questions connected to the deployment and use of RS.
\end{itemize} 

Selected papers will be presented with lightning talks during the workshop.

\section{Format and Duration}

Our workshop will feature a unique interactive event in the form of a hackathon. This will allow participants to collaborate, exchange ideas, and build skills in a supportive and engaging environment. We will provide all the necessary materials and support for the hackathon, and encourage participants to work in teams and submit their projects for consideration.

Aside from the hackathon, we will have a more traditional presentation session where participants can share their research. Additionally, two keynote speakers will share their insights and experiences related to the workshop's topic. We understand that the workshop's value extends beyond the event itself. Therefore, we will release all workshop artifacts, including video recordings and hackathon materials, in an open and accessible format. This will allow participants to revisit and learn from the workshop even after its conclusion, and we hope it will contribute to advancing research in the field.

\subsection{Interactive Activity: Hackathon}
\label{sec:hackathon}

Our hackathon has been organised as a half-day activity, taking place after the presentations of accepted papers. We will ask participants to come up with a contribution for the rounded evaluation of RS, leveraging an agreed-upon dataset, open-source code, and tools prepared in advance by the organizers. Contribution details will be intentionally left open-ended, as we would like participants to engage different angles of the problem on a shared set of resources. Examples could be operationalizing important notions of robustness, applying and discussing metric definitions from literature, quantifying the trade-off between privacy and accuracy, and so on. The hackathon is a unique opportunity to ``live and breathe'' the workshop themes, increase chances of multi-disciplinary collaboration, network and discover related work by peers, and contribute valuable materials back to the community. 

We plan to release materials in advance of the workshop so that participants can prepare. The event will be open to all the attendees, but we will ask them to register in advance to better prepare the event -- for example, we may want to create teams ahead of the event to maximize their diversity.

We plan to award monetary prizes to different contributions in different categories: as the contributions are open-ended, so are the criteria. Generally, we lean towards awarding innovative methodologies and clever ideas, with particular attention to real-world impact: new metrics, interesting model analyses, and thorough qualitative evaluations are all in scope.

\paragraph{Datasets} We will be working in a \textbf{Two-Sided Digital Aggregator} scenario. 
We reconciled tracks from the EvalRS2022 dataset with those in the WASABI dataset \cite{fell:hal-03812106,buffa:hal-03282619}, significantly augmenting the information we had about them. 
WASABI provides, for a portion of the tracks in the EvalRS dataset (coverage is 48\%, for a total of 361218 unique songs), extra features like valence-arousal predictions, emotion and social tags from last.fm, labels determining whether a song is a classic or not, and topic distributions obtained with a LDA topic model (see \cite{fell2019love} for an in-depth description of these features). Furthermore, we have used Sentence-Bert~\cite{reimers-2019-sentence-bert} with \texttt{all-mpnet-base-v2} general-purpose model
on WASABI song lyrics to enrich our dataset with song embeddings, calculated both on the full lyrics and on individual verses.
The final result is a dataset with rich metadata and strong baselines for models and tests, ideal to investigate both sides of RS in users and items.

\subsection{Sponsorship}

\subsubsection{Workshop}
mozilla.ai will provide sponsorship for the event: the funds will be used to award prizes for important and original work, support students and practitioners from unprivileged backgrounds, and help with the relevant travel expenses (speakers, organizers, etc.).

\subsubsection{Social event}
Bauplan, Snap, Costanoa Venture will generously provide support for the prize ceremony and the social gathering that will happen after the workshop. For up-to-date details on the logistics, check the workshop website regularly.

\section{Tentative Program}

\subsection{Important Dates}

\begin{itemize}
    \item Paper submissions due: June 16th, 2023
    \item Paper acceptance notification: June 23th, 2023
    \item Camera ready deadline: July 6th, 2023
    \item Workshop day: August 7th, 2023
\end{itemize}

Participants should refer to the official website (\url{https://reclist.io/kdd2023-cup/}) for the latest logistic information on the workshop.

\subsection{Keynote Speakers and Program Committee}

To reflect a broader view on RS, we are putting together a diverse set of practitioners from \textbf{industry} and \textbf{academia} as our program committee.

Invited speakers will be chosen from a shortlist of world-renown experts: given the focus on real-world impact, our keynote choice will mostly reflect the importance of ethically relevant deployment scenarios (e.g. misinformation, social networks, biases etc.).

\noindent \textbf{Schedule.} Table~\ref{table:schedule} summarizes our plan for the workshop. Our proposal features a mix of standard event and social gatherings: in particular, our 4 hour workshop during normal KDD hours will be a mix of talks and hackathon (first part). We will then invite participants to stay for a pizza night sponsored by the organizers, and continue the hacking for few more hours: we will then celebrate the winners of the contest at the end of the evening, in a fun and informal environment.

\begin{table}[]
\small
\begin{tabular}{l
>{\columncolor[HTML]{D9EAD3}}l l}
\cellcolor[HTML]{C9DAF8}{\color[HTML]{1D1C1D} \textbf{Time}} & \cellcolor[HTML]{C9DAF8}{\color[HTML]{1D1C1D} \textbf{Activity}} & \cellcolor[HTML]{C9DAF8}{\color[HTML]{1D1C1D} \textbf{Details}}                                                                        \\ \hline
{\color[HTML]{1D1C1D} 30 Minutes}                            & {\color[HTML]{1D1C1D} \textbf{1st Keynote}}                          & {\color[HTML]{1D1C1D} First keynote: Luca Belli}                                                                                                      \\ \hline
{\color[HTML]{1D1C1D} 30 Minutes}                            & {\color[HTML]{1D1C1D} \textbf{Papers}}                           & {\color[HTML]{1D1C1D} Poster presentations OR lighting talks}                                                                                                                                                 \\ \hline
{\color[HTML]{1D1C1D} 2:30 Hours}                               & {\color[HTML]{1D1C1D} \textbf{Hackathon: part I}}                        & {\color[HTML]{1D1C1D} \textit{\begin{tabular}[c] {@{}l@{}}\textbf{10 minutes}: materials, rules and the goals. \\ \\ \textbf{140 minutes}: teamwork\end{tabular}}} \\ \hline
{\color[HTML]{1D1C1D} 30 Minutes}                            & {\color[HTML]{1D1C1D} \textbf{2nd Keynote}}                          & {\color[HTML]{1D1C1D} Second keynote: Joey Robinson}                                                \\ \hline
{\color[HTML]{1D1C1D} Break}                           & {\color[HTML]{1D1C1D} \textbf{Break}}                       & {\color[HTML]{1D1C1D} Break before the pizza night}  \\ \hline  
{\color[HTML]{1D1C1D} 2 Hours}                           & {\color[HTML]{1D1C1D} \textbf{After-party, Hackathon: part II}}                       & {\color[HTML]{1D1C1D} Pizza night: teamwork, project presentations, prizes}  \\ \hline  

\end{tabular}
\caption{Schedule for EvalRS2023}
\label{table:schedule}
\vspace{-2em}
\end{table}

\section{Organizing team}

Workshop organizers are a mix of academic and industry practitioners, with broad experience in RS and their evaluation. Our team includes the main authors of three popular open-source RS packages for training and evaluation (Merlin\footnote{\url{https://github.com/NVIDIA-Merlin/Merlin}}, Elliot~\cite{DBLP:conf/sigir/AnelliBFMMPDN21}, RecList~\cite{DBLP:conf/www/ChiaTBHK22}), as well as veterans in the space of Data Challenges and Open Datasets for recommendations (SIGIR 2021 Data Challenge\footnote{\url{https://sigir-ecom.github.io/ecom2021/data-task.html}}, EvalRS 2022\footnote{\url{https://reclist.io/cikm2022-cup/}}~\cite{DBLP:conf/cikm/TagliabueBSAGMC22}). 
\newline

\noindent \textbf{Federico Bianchi} (Stanford) is a postdoctoral researcher at Stanford University. His research, ranging from NLP methods for textual analytics to recommender systems for e-commerce has been published at major NLP and AI conferences (EACL, NAACL, EMNLP, ACL, AAAI, ICLR, RecSys) and journals (Cognitive Science, Applied Intelligence, Semantic Web Journals). Federico co-organized the SIGIR2021 Data Challenge, and the CIKM2022 Data Challenge.

\noindent \textbf{Patrick John Chia} (Coveo) is an Applied Scientist at Coveo, focusing on research at the intersection of IR, NLP and eCommerce. He is a co-organizer of multiple data challenges (CIKM 2022, SIGIR 2021) and is a speaker and published author at multiple academic and industry venues (ACL, SIGIR, Berlin Buzzwords). His broad interests lie in better understanding AI methods of today and developing AI with more human-like learning capabilities.

\noindent \textbf{Jacopo Tagliabue} (New York University) is co-creator of \textit{RecList}. He is Adj. Professor of MLSys at NYU, speaks regularly at top-tier conferences (including NAACL, WWW, RecSys, SIGIR, KDD), and has served as a committee member for ECNLP, ECONLP, EMNLP, SIRIP, ACL. Jacopo is co-organizer of SIGIR eCom, and was the lead organizer of the SIGIR Data Challenge in 2021, and the CIKM Data Challenge in 2022.

\noindent \textbf{Ciro Greco} (Bauplan) holds a Ph.D. in Linguistics and Cognitive Neuroscience at Milano-Bicocca and was a post-doctoral fellow at Ghent University. In 2017, he founded Tooso, which was acquired in 2019 by Coveo. He published extensively in top-tier conferences (including NAACL, ACL, RecSys, SIGIR) and scientific journals (Cognitive Science, Nature Communications). He was also co-organizer of the SIGIR DC 2021 and the CIKM DC in 2022.

\noindent \textbf{Claudio Pomo} (Politecnico di Bari) is a research fellow at Politecnico di Bari. His research concerns the issues of responsible AI for personalization, with a particular interest in results' reproducibility and multi-objective performance evaluation. Contributions on these topics have been accepted in influential area conferences (SIGIR, RecSys, ECIR, UMAP) and journals (Information Science, IPM). Claudio is among the authors and contributors of Elliot.

\noindent \textbf{Gabriel Moreira} (NVIDIA) holds a Ph.D on RS at ITA, Brazil. He is a Sr. Research Scientist and Engineer at NVIDIA Merlin team. He is recognized as a Google Developer Expert (GDE) for Machine Learning since 2019, was a co-organizer of the CIKM Data Challenge 2022, committee member for RecSys and SIGIR conferences and is a distinguished reviewer for ACM TORS journal.  He was a member of the teams that won recent RS competitions.

\noindent \textbf{Davide Eynard} (mozilla.ai) is a Staff MLE and researcher at mozilla.ai, previously Twitter and Fabula AI. He has been a researcher and lecturer at Università della Svizzera Italiana (USI) and Politecnico di Milano. His research deals with knowledge representation, large-scale multimedia retrieval, computer vision, graph learning, and federated social networks. He has served as a committee member of SIGGRAPH, ECCV, ICCV, CVPR, ICML, NeurIPS, ICLR.

\noindent \textbf{Fahd Husain} (mozilla.ai) is the Director of ML at mozilla.ai. He has worked on DARPA research efforts to counter human trafficking, identify narrative propagation on social media, and navigate biomedical graphs for pandemic research. Most recently, he was Principal Investigator for DARPA's World Modelers and Automated Scientific Knowledge Extraction programs. His research is on graph machine learning, weak supervision, and human-in-the-loop paradigms.

 \section*{Acknowledgments}
FB is supported by the Hoffman–Yee Research Grants Program and the
Stanford Institute for Human-Centered Artificial Intelligence.

\newpage
\bibliographystyle{ACM-Reference-Format}
\bibliography{biblio}

\end{document}